\documentclass[a4paper,11pt]{article}
\usepackage{pos}

\title{The Evolution of Lattice Field Theory: a Statistical Study}

\author*{Wolfgang Bietenholz}

\affiliation[]{Instituto de Ciencias Nucleares \\
  Universidad Nacional Aut\'{o}noma de M\'{e}xico \\
  A.P.\ 70-542, C.P.\ 04510 Ciudad de M\'{e}xico, Mexico}

\emailAdd{wolbi@nucleares.unam.mx}

\abstract{Researchers working in lattice field theory constitute an
established community since the early 1990s, and around the same
time the online open-access e-print repository {\tt arXiv} was
created. The fact that this field has a specific
arXiv section, {\tt hep-lat}, provides a unique opportunity for a
statistical study of its evolution over the last three decades.
We present data for the number of entries, $E$, published papers, $P$,
and citations, $C$, in total and separated by nations. We compare them
to 6 other {\tt arXiv} sections,
and to socio-economic indices of the nations involved, namely
the Gross Domestic Product (GDP) and the Education Index (EI).
We present rankings, which are based either on the Hirsch Index $H$,
or on the linear combination $\Sigma = E + P + 0.05 C$. We consider
both extensive and intensive national statistics, {\it i.e.}\
absolute and relative to the population or to the GDP.}

\FullConference{%
 The 38th International Symposium on Lattice Field Theory, LATTICE2021
  26th-30th July, 2021
  Zoom/Gather@Massachusetts Institute of Technology}

\begin{document}
\maketitle

\section{Outline}
\vspace*{-2mm}

The conceptual basis of lattice field theory was elaborated in the
1970s and 1980s, and since the early 1990s physicists working in this
field are a well-established, intercontinental community. The latter
is related to the fact that around the same time computational
resources became accessible more easily --- computing was
``democratized'' --- and they attained a level which allows for
precise non-perturbative studies of some quantum field theoretic models,
by means of Monte Carlo simulations in the lattice regularization.

It happened around the same time, more precisely in 1991/2, that the
online e-print repository {\tt arXiv} \cite{arXiv}
became operational. Ever since it has
contributed very significantly to the scientific communication.
In particular, since 1992 its {\tt hep-lat} section captures practically
the entire activity of the lattice community.\footnote{Also since 1992,
  the annual Lattice Conference (or Symposium) attained an extent of
  $\gtrsim 200$ contributions.}
Lattice researchers
use it comprehensively, which is not the case to the same extent
in other branches of physics, like condensed matter and optics.

The {\tt hep-lat} section provides a perfect opportunity to
statistically monitor the evolution of lattice field theory over
the last three decades.
In contrast, a similar study in other specific lines of physical
research would require a tedious and less reliable search for
keywords in titles and abstracts.

Taking advantage of this opportunity, we study the number of articles
which were submitted to the {\tt arXiv} with {\tt hep-lat} as the
``primary archive'', {\tt primarch} (we do not consider articles
which are cross-listed to {\tt hep-lat}; usually they only have an
indirect link to lattice field theory).
We used the open access digital library {\tt INSPIRE} \cite{INSPIRE}
to count the following quantities:
\begin{table}[h!]
\centering
\begin{tabular}{|c|l|}
\hline
$E$ & Entries, all article submitted to the {\tt arXiv} with
primary archive {\tt hep-lat} \\
\hline
$P$ & The subset of $E$, which was later published as regular
papers (this excludes proceeding \\
& contributions and unpublished preprints) \\
\hline
$C$ & Citations to all articles in $E$, which were registered by
{\tt INSPIRE} until summer 2020 \\
\hline
$H$ &  Hirsch Index \cite{Hirsch} of a set of articles which appear
in $E$, considering all the citations to them \\
& included in $C$.\\
\hline
\end{tabular}
\end{table}

In order to compare lattice field theory to other fields of physical
research, we considered the same quantities in 6 other arXiv sections,
which are thematically somewhat related, namely
{\tt hep-ph} (high-energy physics, phenomenology),
{\tt hep-th} (high-energy physics, theory),
{\tt gr-qc} (general relativity and quantum cosmology),
{\tt nucl-th} (nuclear theory),
{\tt quant-ph} (quantum physics),
and {\tt cond-mat$\star$} (condensed matter;
the symbol $\star$ indicated the sum over all subsections).

In the national statistics,
an article counts for a nation if at least one author has a working
address there (we are not concerned with the authors' nationalities).
Hence it can count for several countries.
The data were taken in July 2020 from the ``old'' INSPIRE version,
{\tt https://old.inspirehep.net},
which is deactivated now, unfortunately.\footnote{The new {\tt INSPIRE}
  version \cite{INSPIRE} is not useful for such a study: if we fix
  {\it e.g.}\ ``date'', much larger numbers appear than in the old version,
  due to multiple counting. The command ``de'' (data earliest)
  is supposed to overcome this effect, but it returns weird results, which
  are much too low ({\it e.g.}\ just one {\tt hep-lat} entry in 2019).
  Moreover, the new version does not have the option ``country code'',
  hence it does not provide data for national statistics.}

We are going to present data for {\em global} and for {\em national}
statistics, first from an {\em extensive} and then from an
{\em intensive} perspective, {\it i.e.}\ absolute and relative to
the population, respectively. (We do not count contributions by
specific authors or collaborations.)
In the special case of Switzerland, we exclude CERN as an affiliation.
As a single parameter, the Hirsch Index $H$ \cite{Hirsch} can be applied
in extensive rankings, but not in intensive ones. As another single
parameter for the scientific activity and achievement, we define
\begin{equation}  \label{Sigmadef}
  \Sigma := E + P + 0.05 C \ ,
\end{equation}
where the coefficients (``weights'') are motivated by the statistical
trends to be presented below, see in particular Table \ref{tabarXivs}.
The index $\Sigma$ does have an intensive counterpart, see Section 3.

We compare these data with two socio-economic parameters: as an economic index,
we consider the Gross Domestic Product (GPD), given in $10^{9}$ US
dollars (more precisely: the value of its purchasing parity in 2011).
Its intensive version is the GPDpp (per capita).
As another intensive quantity, which seems likely to be related to
the scientific performance, we consider the Education Index
(EI),\footnote{In another statistical study, which deals with High Energy
  Physics in Latin America \cite{hepLA}, we considered separately the EI and
  the Human Development Index (HDI). The HDI is the geometric mean of indices
  of income (in purchasing parity), health and the EI. However, the
  EI and HDI tend to be similar in most countries, so here we refrain
  from a separate consideration.}
which is defined as
\begin{equation}
{\rm EI} := \frac{1}{2} \Big( \frac{\rm EYS}{18} + \frac{\rm MYS}{15} \Big) \ . 
\end{equation}
EYS means ``expected years of schooling'' for children (normalized by
the duration for a Master's degree), while MYS are the
``mean years of schooling'' of adults
(normalized to the projected maximum in 2025). For each nation and year
one obtains ${\rm EI} \in [0, 1]$. Our source for the annual GDP,
GPDpp and EI is the United Nations Development Programme \cite{UN}; we
average over the years from 1992 to 2019. The population is averaged
over the same period; it is expressed in millions of inhabitants
(we do not include tiny countries with less than $10^{5}$ inhabitants).
We also intended to include the percentage of
``skilled labor force'', but the data for different countries do not
seem to be based on consistent criteria.\footnote{For instance, Japan
  reports 99.9\% ``skilled labor force'', but Italy and Portugal only
  69.6\% and 54.1\%, respectively.}

\vspace*{-2mm}
\section{Extensive Statistics}
\vspace*{-2mm}

\begin{figure}[h!]
\vspace*{-4mm}
\begin{center}
\includegraphics[angle=270,width=.5\linewidth]{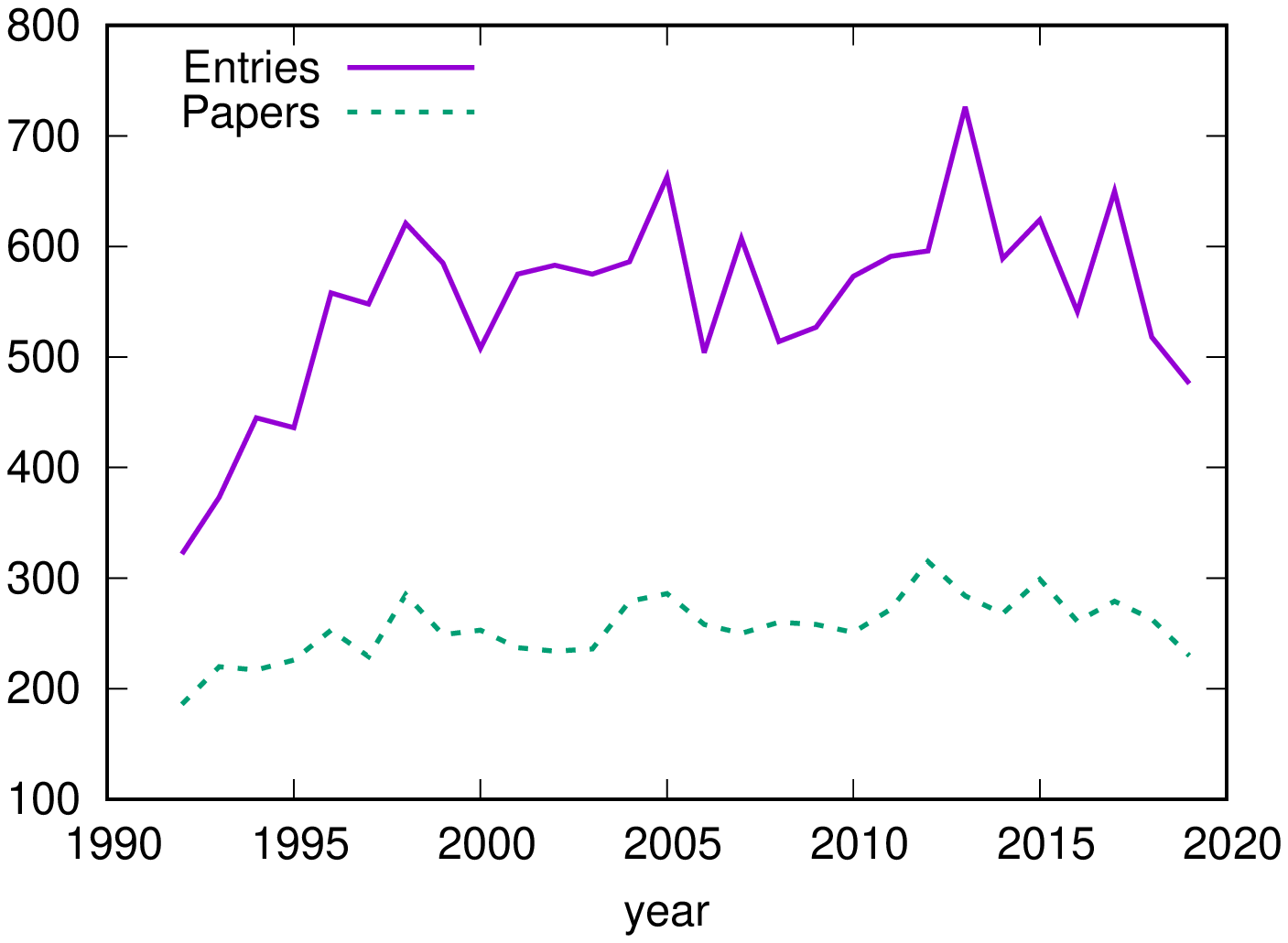}  
  \hspace*{-3mm}
\includegraphics[angle=270,width=.5\linewidth]{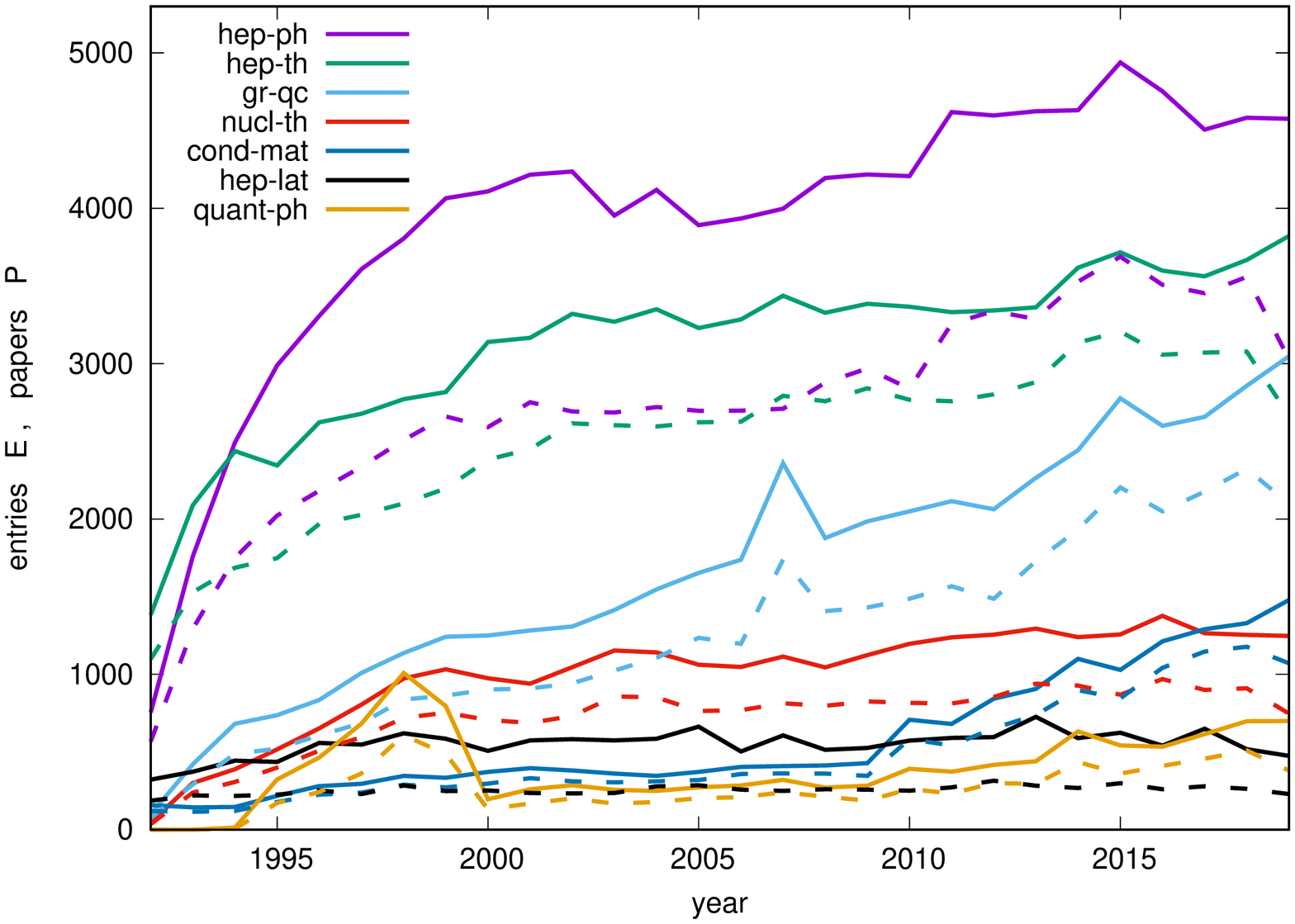}
\end{center}
\vspace*{-5mm}
\caption{Research activity from 1992 to 2019. Left:
  {\tt hep-lat} entries ($E$) and papers ($P$).
  Right: $E$ [solid] and $P$ [dashed] in the {\tt arXiv} sections {\tt hep-ph}
  (purple), {\tt hep-th} (green), {\tt gr-qc} (cyan), {\tt nucl-th} (red),
  {\tt cond-mat} (blue), {\tt hep-lat} (black) and {\tt quant-ph} (orange).
  They are hierarchically ordered, cf.\ Table \ref{tabarXivs}. We see
  some peaks of $E$ and $P$, in particular for {\tt quant-ph} in the
  period from 1995 to 2000. Condensed matter physics has a high productivity,
  but only part of its articles are submitted to the {\tt arXiv};
  we see, however, that this is becoming more frequent since 2008.}
\label{heplatevo}
\end{figure}

Figure \ref{heplatevo} shows global data for the annual evolution
from 1992 to 2019 in the {\tt hep-lat}, and in all the 7
{\tt arXiv} sections under consideration
(in 1991 there were only a few sporadic entries and 2020
we could only capture the first half). The total statistics,
summed up from 1991 to July 2020, for each of the 7 arXiv sections
is displayed in Table \ref{tabarXivs}.

\begin{figure}[h!]
\vspace*{-2mm}
\begin{center}
\includegraphics[angle=270,width=.5\linewidth]{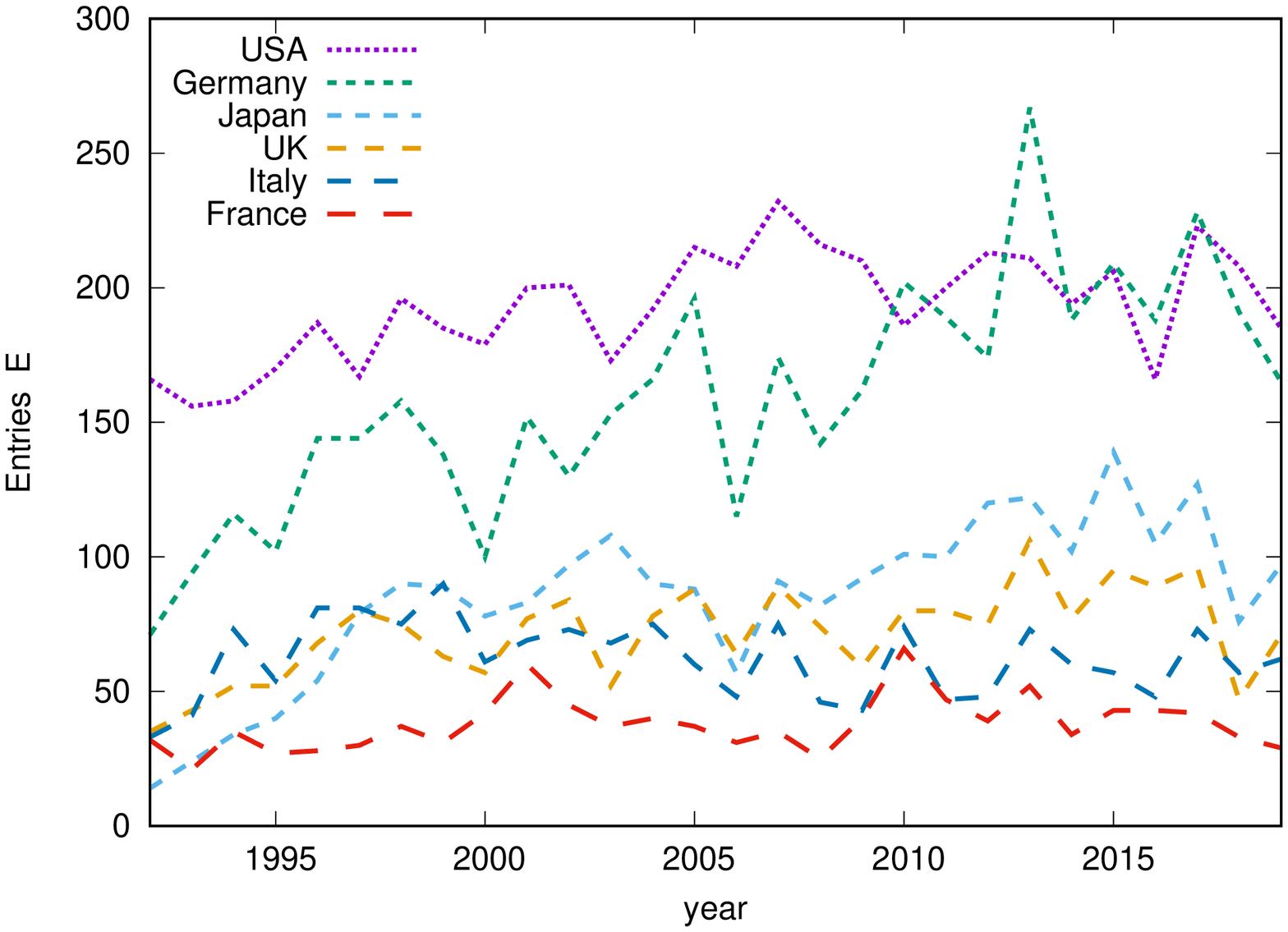}  
  \hspace*{-3mm}
\includegraphics[angle=270,width=.5\linewidth]{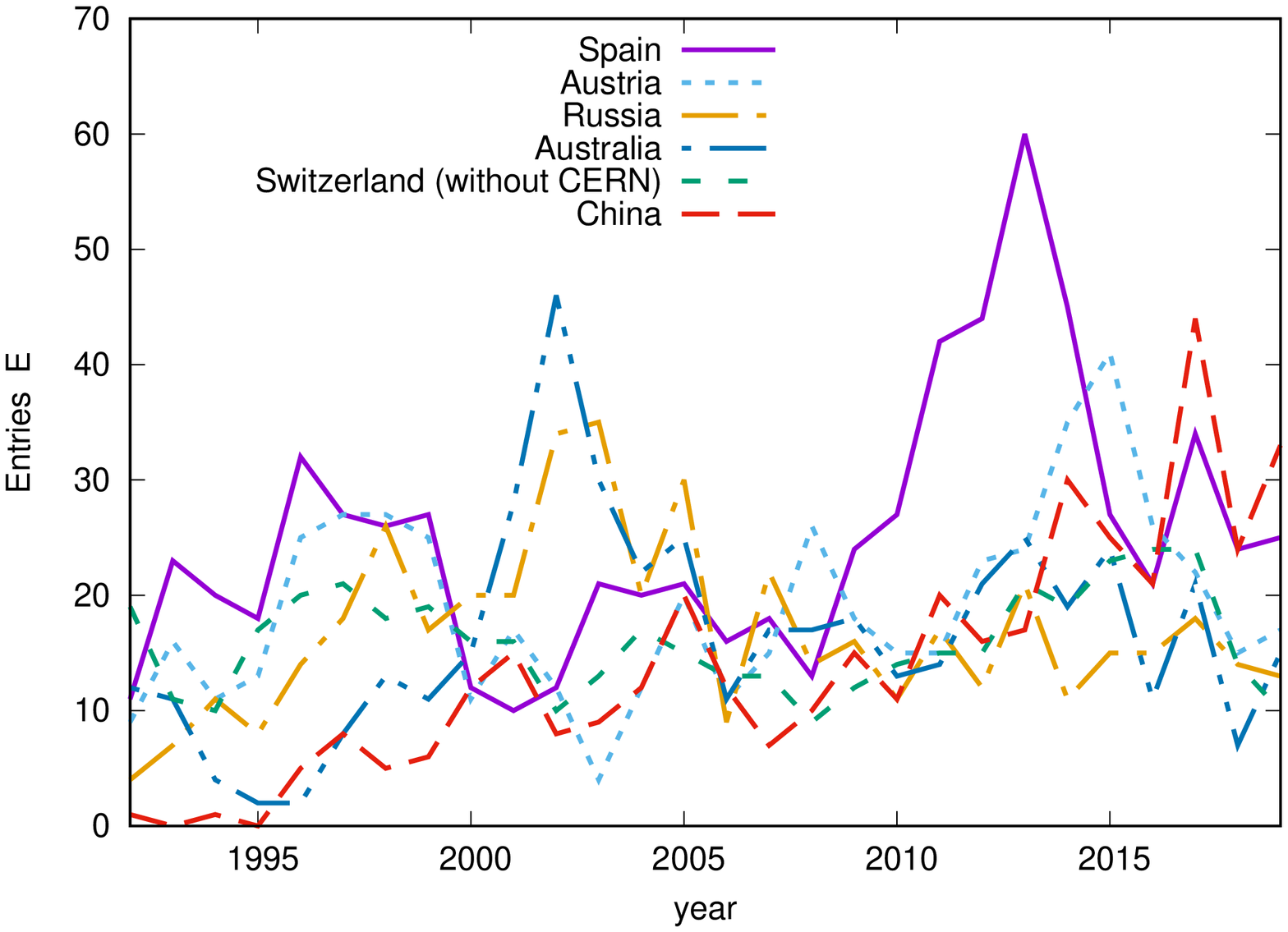}
\end{center}
\vspace*{-5mm}
\caption{Evolution of the number of \,{\tt hep-lat} entries ($E$) of the 6
  leading nations (left), and of the nations ranking $7$ to $12$ (right),
  cf.\ Table \ref{heplatrankingext}. We see a slight general trend up.
During the last decade, Germany has caught up with the USA as the most
productive countries, followed by Japan, UK, Italy, France, etc.
There are remarkable peaks for Germany, Spain, Austria and Australia.
Lately China is moving up.}
\label{Elat1to16}
\end{figure}

Figure \ref{Elat1to16} shows the individual time evolutions of
{\tt hep-lat}-entries $E$ for the 12 leading nations in this respect,
with $E>400$. Table \ref{heplatrankingext} contains the national
sums of $E$, $P$ and $C$ from 1991 to July 2020, along with the population
in millions and the GDP (both averaged from 1992 to 2019). For
comparison, we add the European Union (with 28 nations,
still including the UK) and CERN, but the ranking only
refers to nations. It is based on the Hirsch Index, $H$-rank,
but we also show the $\Sigma$-rank, which is very similar.
This confirms that the definition of $\Sigma$ of eq.\ (\ref{Sigmadef})
is sensible.

\begin{table}
\centering
\begin{tabular}{|c||r|r|r|r|r|r|r|}
\hline
 & {\tt hep-ph} & {\tt hep-th} & {\tt gr-qc} & {\tt nucl-th}
 & {\tt cond-mat} & {\tt hep-lat} & {\tt quant-ph} \\
\hline
$E$ &  111515 &   89279 &   48927 &  28522 &  16969 &   15610 & 11602 \\
$P$ &   76520 &   70561 &   35703 &  20215 &  13677 &    7165 &  7484 \\
$C$ & 3960720 & 2857462 & 1043823 & 682874 & 247734  & 402121 & 106901 \\
\hline
$P/E$ & 0.69 & 0.79 & 0.71 & 0.71 & 0.81 & 0.46 & 0.65 \\
$C/E$ & 35.5 & 32.0 & 21.3 & 23.9 & 14.6 & 25.8 & 9.21 \\
\hline
\end{tabular}
\caption{Total parameters $E$, $P$ and $C$, for 7 arXiv sections,
  summed up from 1991 to July 2020, along with the publication fraction
  $P/E$ and the citation rate $C/E$. Note the peculiarity $P < E/2$
  for {\tt hep-lat}, which might be related to the particular importance
  of the proceedings of the annual Lattice Conference; for a number of
  results, the authors are satisfied if the appear in these ``lattice
  proceedings''. Also the citation rate varies strongly between the
  different sections.}
\label{tabarXivs}
\vspace*{-1mm}
\end{table}

\begin{table}[h!]
\centering
\vspace*{-7mm}
{\small
\begin{tabular}{|l c||r|r||r|r|r||c|r|}
\hline
& $\Sigma$-rank & population & GDP~ & $E$~~ & $P$~~ & $C$~~~ & $\Sigma$ & $H$ \\
\hline
~~~~ European Union
               &    & 495.0 & 15900.4 & 9020 & 4265 & 249236 & {\bf 25746.8} & {\bf 182} \\
~1. USA        &  1 & 294.3 & 14050.1 & 5472 & 2565 & 175639 & {\bf 16818.9} & {\bf 167} \\
~2. Germany    &  2 &  81.3 &  3167.4 & 4523 & 2107 & 143667 & {\bf 13813.3} & {\bf 155} \\
~3. UK         &  3 &  61.4 &  2136.2 & 2026 &  950 &  67386 & {\bf  6345.3} & {\bf 121} \\
~4. Japan      &  4 & 127.6 &  4458.1 & 2419 & 1078 &  55729 & {\bf  6283.5} & {\bf 103} \\
~5. France     &  6 &  61.1 &  2232.9 & 1070 &  540 &  42060 & {\bf  3713.0} & {\bf 101} \\  
~6. Italy      &  5 &  58.5 &  2066.6 & 1765 &  877 &  41252 & {\bf  4704.6} & {\bf  89} \\
~~~~ CERN      &    &       &         &  615 &  306 &  27553 & {\bf  2298.7} & {\bf  79} \\
~7. Switzerland&  8 &   7.5 &   400.7 &  451 &  228 &  16744 & {\bf  1516.2} & {\bf  75} \\
~8. Spain      &  7 &  43.6 &  1326.8 &  708 &  370 &  19310 & {\bf  2043.5} & {\bf  71} \\
~9. Australia  &  9 &  18.9 &   801.9 &  471 &  250 &  14208 & {\bf  1431.4} & {\bf  64} \\
10. Hungary    & 10 &  10.1 &   208.5 &  292 &  151 &  18824 & {\bf  1384.2} & {\bf  61} \\
11. Cyprus     & 14 &   1.0 &    23.1 &  389 &  193 &   8794 & {\bf  1021.7} & {\bf  54} \\  
12. Austria    & 11 &   8.3 &   333.8 &  534 &  254 &  10300 & {\bf  1303.0} & {\bf  51} \\
13. China      & 13 &1326.5 &  9638.8 &  415 &  223 &   9708 & {\bf  1123.4} & {\bf  50} \\
14. Denmark    & 15 &   5.4 &   233.1 &  322 &  174 &  10188 & {\bf  1005.4} & {\bf  50} \\
15. Russia     & 12 & 145.5 &  2822.8 &  475 &  252 &   8608 & {\bf  1157.4} & {\bf  48} \\
16. Canada     & 17 &  32.5 &  1269.4 &  325 &  164 &   8387 & {\bf   908.3} & {\bf  46} \\
17. India      & 16 &1142.0 &  4509.3 &  316 &  183 &   8198 & {\bf   908.9} & {\bf  45} \\
18. Ireland    & 18 &   4.2 &   183.7 &  292 &  117 &   8565 & {\bf   837.3} & {\bf  45} \\
19. Taiwan     & 20 &  22.5 &   702.2 &  196 &  100 &   5491 & {\bf   570.6} & {\bf  40} \\
20. Finland    & 23 &   5.3 &   191.6 &  165 &   81 &   4460 & {\bf   469.0} & {\bf  36} \\
21. Netherlands& 19 &  16.3 &   696.5 &  203 &   98 &   4544 & {\bf   528.2} & {\bf  35} \\ 
22. Israel     & 22 &   6.6 &   197.8 &  149 &   96 &   4622 & {\bf   476.1} & {\bf  35} \\
23. South Korea& 21 &  48.3 &  1242.6 &  238 &   82 &   4448 & {\bf   542.4} & {\bf  34} \\
24. Poland     & 24 &  38.3 &   696.8 &  181 &   82 &   3192 & {\bf   422.6} & {\bf  30} \\
25. Brazil     & 25 & 184.4 &  2397.1 &  117 &   58 &   3939 & {\bf   372.0} & {\bf  29} \\
26. Sweden     & 27 &   9.2 &   365.9 &   99 &   50 &   2162 & {\bf   257.1} & {\bf  27} \\
27. Slovenia   & 30 &   2.0 &    51.1 &   67 &   33 &   2174 & {\bf   208.7} & {\bf  26} \\ 
28. Portugal   & 26 &  10.4 &   264.0 &  138 &   55 &   2252 & {\bf   305.6} & {\bf  25} \\
29. Slovakia   & 28 &   5.4 &   111.5 &   82 &   42 &   2354 & {\bf   241.7} & {\bf  25} \\
30. Greece     & 29 &  10.9 &   273.9 &   96 &   49 &   1756 & {\bf   232.8} & {\bf  22} \\
31. Belgium    & 31 &  10.7 &   409.5 &   54 &   28 &   1792 & {\bf   171.6} & {\bf  20} \\   
32. Mexico     & 33 & 106.6 &  1710.8 &   58 &   26 &    760 & {\bf   122.0} & {\bf  15} \\
33. Ukraine    & 32 &  47.5 &   319.9 &   85 &   34 &    417 & {\bf   139.9} & {\bf  11} \\
34. Turkey     & 34 &  68.2 &  1192.0 &   25 &   19 &    347 & {\bf    61.3} & {\bf  10} \\
35. New Zealand& 36 &   4.1 &   127.7 &   11 &    8 &    420 & {\bf    40.0} & {\bf 8} \\ 
36. Bangladesh & 39 & 137.0 &   317.0 &   10 &    6 &    267 & {\bf   29.4} & {\bf 7} \\
37. Belarus    & 35 &   9.7 &   113.6 &    9 &    6 &    527 & {\bf   41.3} & {\bf 6} \\
38. Iran       & 37 &  69.9 &  1088.1 &   17 &   11 &    126 & {\bf   34.3} & {\bf 6} \\
39. Georgia    & 38 &   4.4 &    22.6 &   11 &    7 &    236 & {\bf   29.8} & {\bf 5} \\
40. Norway     & 40 &   4.7 &   280.4 &   14 &    6 &     83 & {\bf   24.2} & {\bf 5} \\
41. Albania     & 41 &   3.0 &    22.9 &   15 &    2 &    107 & {\bf 22.4} & {\bf 5} \\
42. Singapore   & 43 &   4.5 &   297.6 &   10 &    5 &    113 & {\bf 20.7} & {\bf 4} \\
\hline
\end{tabular}
}
\vspace*{-1.5mm}
\caption{Extensive {\tt hep-lat} statistics:
  ranking according to the Hirsch Index $H$,
  which is very similar to the $\Sigma$-rank.
  Both ranks refer to nations only;
  in case of identical $H$-indices, $\Sigma$ decides;
  further nations with $H=4$ are Croatia, Uruguay and the Czech Republic.
  According to the $\Sigma$-rank, Thailand is at position 42.
  Population in millions of inhabitants, Gross Domestic
  Product (GDP) in $10^9$ US \$ (purchasing parity in 2011),
  both averaged from 1992 to 2019.
  We add the European Union --- with 28 nations, still including
  the UK --- and CERN. Here and throughout,
  the data for Switzerland exclude CERN.}
\label{heplatrankingext}
\end{table}

\begin{table}
\centering
\vspace*{-7mm}
{\small
\begin{tabular}{|l c||r|r|r||r|r|}
  \hline
& $\Sigma$-rank & $E$~~ & $P$~~ & $C$~~~ & $\Sigma$~~~~~ & $H$ \\
\hline
~1. USA         &  1 &  85601 &  63422 & 4006414 & {\bf 349343.7} & {\bf 583} \\
~~~~ European
Union           &    & 147464 & 106804 & 4961069 & {\bf 502321.4} & {\bf 548} \\
~2. France      &  3 &  28598 &  21090 & 1368170 & {\bf 118096.5} & {\bf 394} \\ 
~3. Germany     &  2 &  45242 &  35533 & 1713406 & {\bf 166445.3} & {\bf 390} \\
~4. UK          &  4 &  26174 &  19927 & 1089933 & {\bf 100597.7} & {\bf 339} \\
~~~~ CERN       &    &   9836 &   7253 &  682730 & {\bf  51225.5} & {\bf 333} \\
~5. Italy       &  5 &  27034 &  19875 &  949801 & {\bf  94399.1} & {\bf 309} \\
~6. Spain       &  9 &  16466 &  12223 &  601111 & {\bf  58744.6} & {\bf 267} \\
~7. Russia      &  7 &  22429 &  15260 &  586236 & {\bf  67000.8} & {\bf 257} \\
~8. Canada      & 10 &  12364 &   9696 &  469849 & {\bf  45552.5} & {\bf 250} \\
~9. Japan       &  6 &  24611 &  18620 &  690235 & {\bf  77742.8} & {\bf 249} \\
10. Switzerland & 14 &  5734  &   4142 &  290056 & {\bf  24378.8} & {\bf 227} \\
11. Netherlands & 16 &   5634 &   4307 &  262786 & {\bf  23080.3} & {\bf 207} \\
12. Poland      & 13 &   8205 &   5660 &  245108 & {\bf  26120.4} & {\bf 191} \\
13. China       &  8 &  20475 &  16372 &  444642 & {\bf  59079.1} & {\bf 186} \\
14. India       & 11 &  14374 &  10925 &  332859 & {\bf  41942.0} & {\bf 186} \\
15. Sweden      & 18 &   4973 &   3753 &  198321 & {\bf  18642.1} & {\bf 178} \\
16. Belgium     & 17 &   4955 &   3798 &  201122 & {\bf  18809.1} & {\bf 171} \\
17. Israel      & 19 &   5085 &   3943 &  176081 & {\bf  17832.1} & {\bf 166} \\
18. Brazil      & 12 &  11460 &   8961 &  222670 & {\bf  31554.5} & {\bf 146} \\
19. South Korea & 15 &   7627 &   6096 &  196954 & {\bf  23570.7} & {\bf 142} \\
20. Portugal    & 22 &   4166 &   3087 &  128890 & {\bf  13697.5} & {\bf 139} \\
21. Austria     & 23 &   3885 &   2563 &  112541 & {\bf  12075.1} & {\bf 137} \\
22. Greece      & 24 &   3505 &   2748 &  107266 & {\bf  11616.3} & {\bf 134} \\
23. Australia   & 20 &   4185 &   3219 &  136125 & {\bf  14210.3} & {\bf 133} \\
24. Denmark     & 27 &   3006 &   2248 &  103328 & {\bf  10420.4} & {\bf 133} \\
25. Taiwan      & 21 &   4110 &   3186 &  134243 & {\bf  14008.2} & {\bf 129} \\
26. Finland     & 29 &   2460 &   1818 &   86101 & {\bf   8583.1} & {\bf 128} \\
27. Hungary     & 30 &   2274 &   1586 &   88347 & {\bf   8277.4} & {\bf 121} \\
28. Chile       & 28 &   3089 &   2567 &   77183 & {\bf   9515.2} & {\bf 108} \\
29. Argentina   & 31 &   2576 &   2179 &   63899 & {\bf   7950.0} & {\bf 104} \\
30. Mexico      & 25 &   4450 &   3232 &   75687 & {\bf  11466.4} & {\bf 101} \\
31. Ireland     & 36 &   1476 &   1078 &   43298 & {\bf   4718.9} & {\bf  97} \\
32. Iran        & 26 &   4054 &   3398 &   64526 & {\bf  10678.3} & {\bf  89} \\
33. South Africa & 34 &   1903 &   1483 &   44556 & {\bf 5613.8} & {\bf 85} \\
34. Slovenia     & 41 &    718 &    484 &   28219 & {\bf 2613.0} & {\bf 85} \\
35. Norway       & 37 &   1242 &    923 &   35198 & {\bf 3924.9} & {\bf 84} \\
36. Ukraine      & 33 &   2508 &   1674 &   42411 & {\bf 6302.6} & {\bf 82} \\
37. Czech Rep.   & 35 &   2106 &   1501 &   37754 & {\bf 5494.7} & {\bf 78} \\
38. Bulgaria     & 40 &   1040 &    684 &  24849  & {\bf 2966.5} & {\bf 73} \\
39. Croatia      & 38 &   1035 &    778 &   25216 & {\bf 3073.8} & {\bf 71} \\
40. Estonia      & 47 &    425 &    336 &   20056 & {\bf 1763.8} & {\bf 71} \\
41. Turkey       & 32 &   2571 &   2112 &   32563 & {\bf 6311.2} & {\bf 62} \\
42. Georgia      & 45 &    693 &    512 &   16033 & {\bf 2006.7} & {\bf 62} \\
43. Armenia      & 42 &    835 &    635 &   17331 & {\bf 2336.6} & {\bf 61} \\
44. Slovakia     & 48 &    642 &    394 &   13173 & {\bf 1694.7} & {\bf 61} \\
45. New Zealand  & 49 &    495 &    362 &   14576 & {\bf 1585.8} & {\bf 60} \\
\hline
\end{tabular}
}
\caption{Like Table \ref{heplatrankingext}, but with extensive,
joint statistics for the 7 arXiv sections under consideration,
for all nations with $H \geq 60$.
Based on the $\Sigma$-rank, Romania (39), Colombia (43) and Pakistan (44)
enter the top 45.}
\label{7arxivrankingext}
\end{table}

\vspace*{-2mm}
\section{Intensive Statistics}
\vspace*{-2mm}

\begin{figure}[h!]
\vspace*{-3mm}
\begin{center}
  \includegraphics[angle=270,width=.51\linewidth]{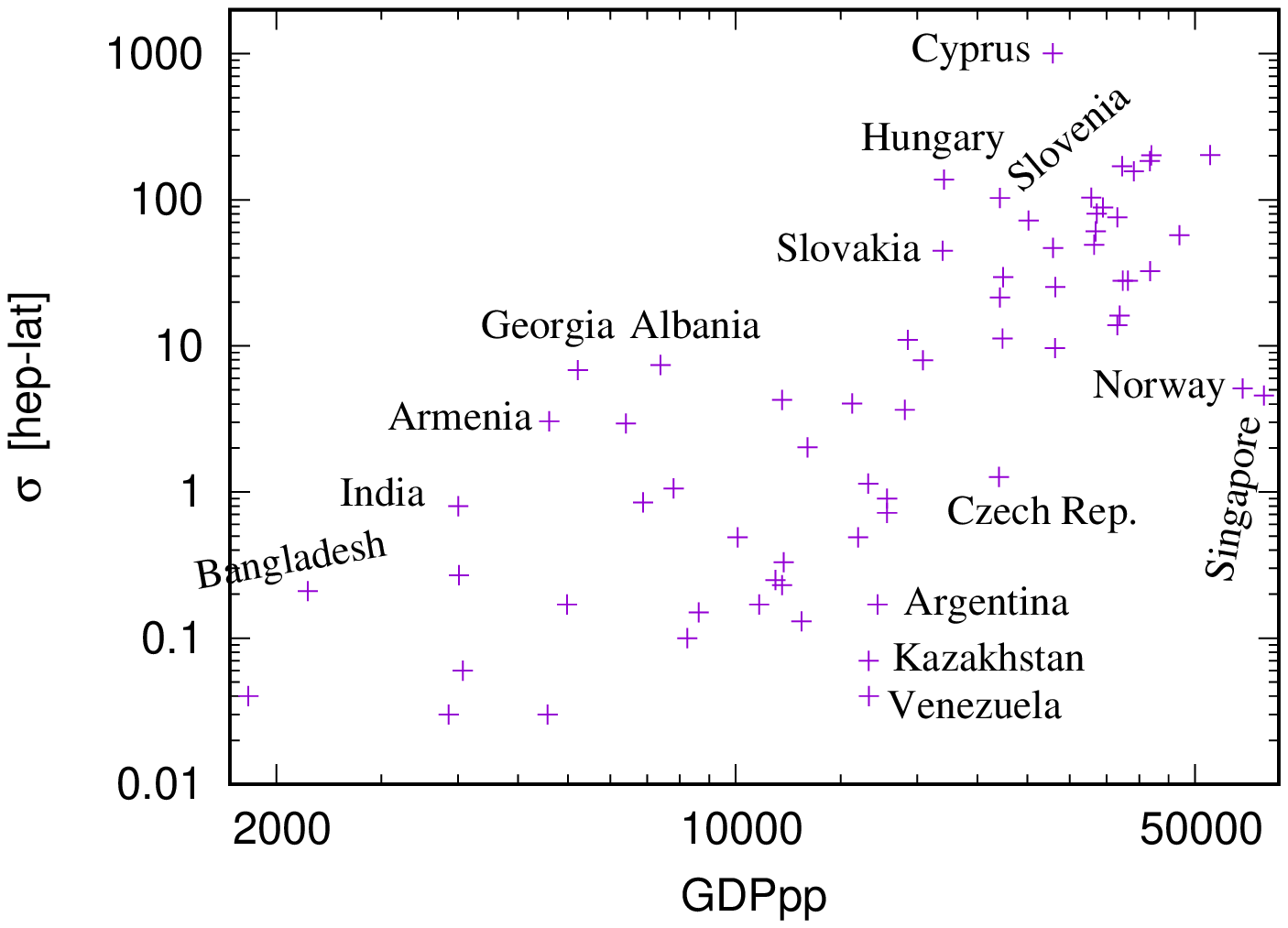}
  \hspace*{-5mm}
  \includegraphics[angle=270,width=.51\linewidth]{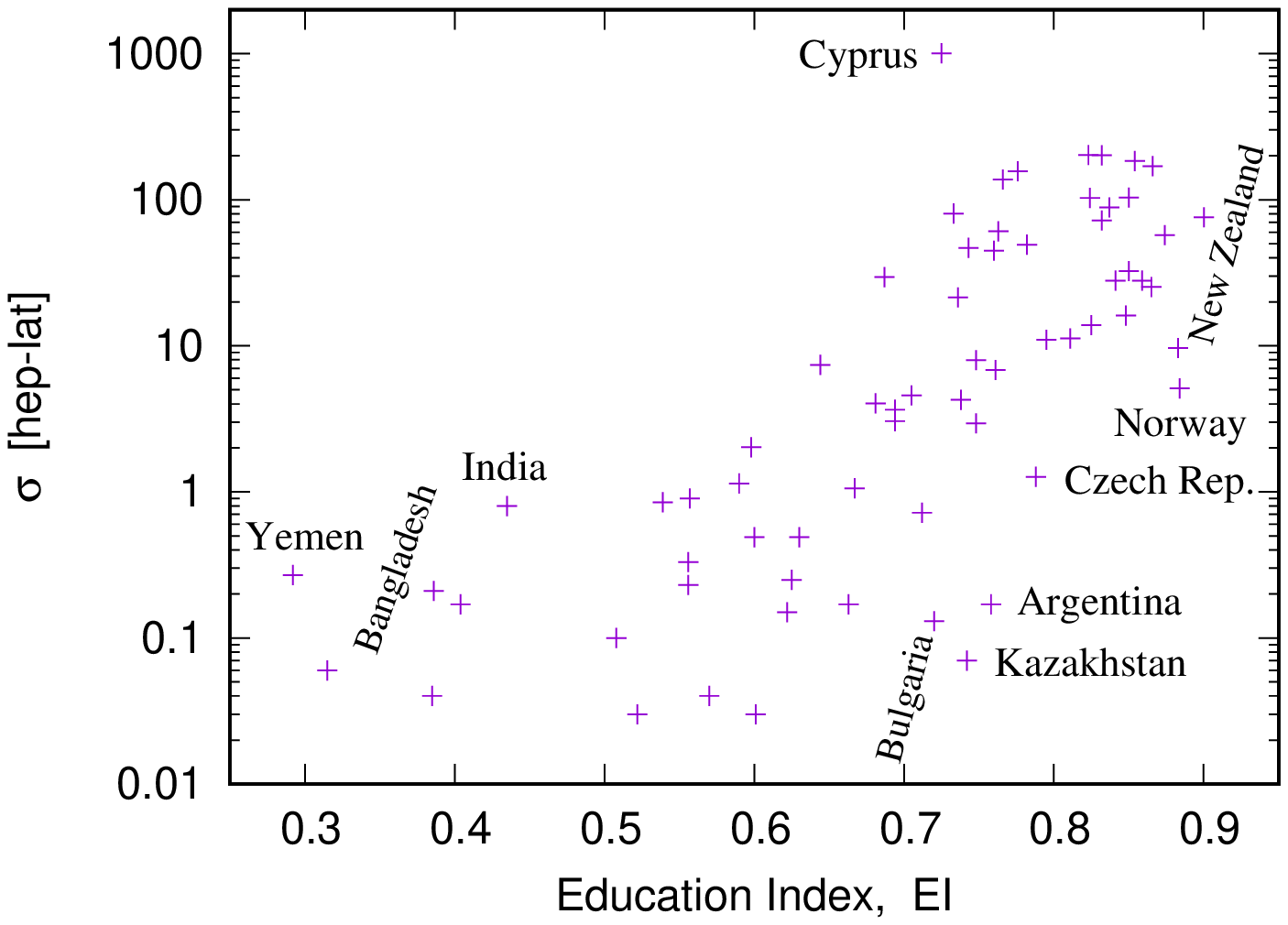} 
\end{center}
\vspace*{-5mm}
\caption{Scatter plots for the 66 nations which contributed any {\tt hep-lat}
  entries.
We show $\sigma := \Sigma /{\rm pop}$
(pop = population in millions) vs.\ GDPpp (left), and vs.\ EI (right).
Monotonic trends are visible, but not as clearly as one might expect.
The top 45 nations can be identified from Table \ref{heplatrankingint}.
We indicate some nations which are clearly off the dominant trend
(above or below).}
\label{scatterheplat}
\end{figure}

Figure \ref{scatterheplat} shows scatter
plots for $\sigma := \Sigma /{\rm pop}$ (pop: population in millions of
inhabitants) vs.\ GPDpp and vs.\ EI. We include the 101 nations with $H>5$.
The symbols for the 45 dominant nations in these plots can
be identified from Tables \ref{heplatrankingint} and \ref{7arxivrankingint}
for the intensive  {\tt hep-lat} statistics and summed over
the 7 {\tt arXiv} sections, respectively.
These tables for intensive quantities further include the GDPpp, the
EI and $(e,p,c) := (E,P,C) /{\rm pop}$. Here the ranking is based
on the parameter $\sigma = \Sigma/{\rm pop}$,
but we also display the economic rank (e-rank) based on
$\Sigma/{\rm GPD} \propto \sigma /{\rm GPDpp}$.
\vspace*{3mm}

{\it This work was supported by UNAM-DGAPA through PAPIIT project IG100219.}

\begin{figure}[h!]
\vspace*{-3mm}
\begin{center}
  \includegraphics[angle=270,width=.51\linewidth]{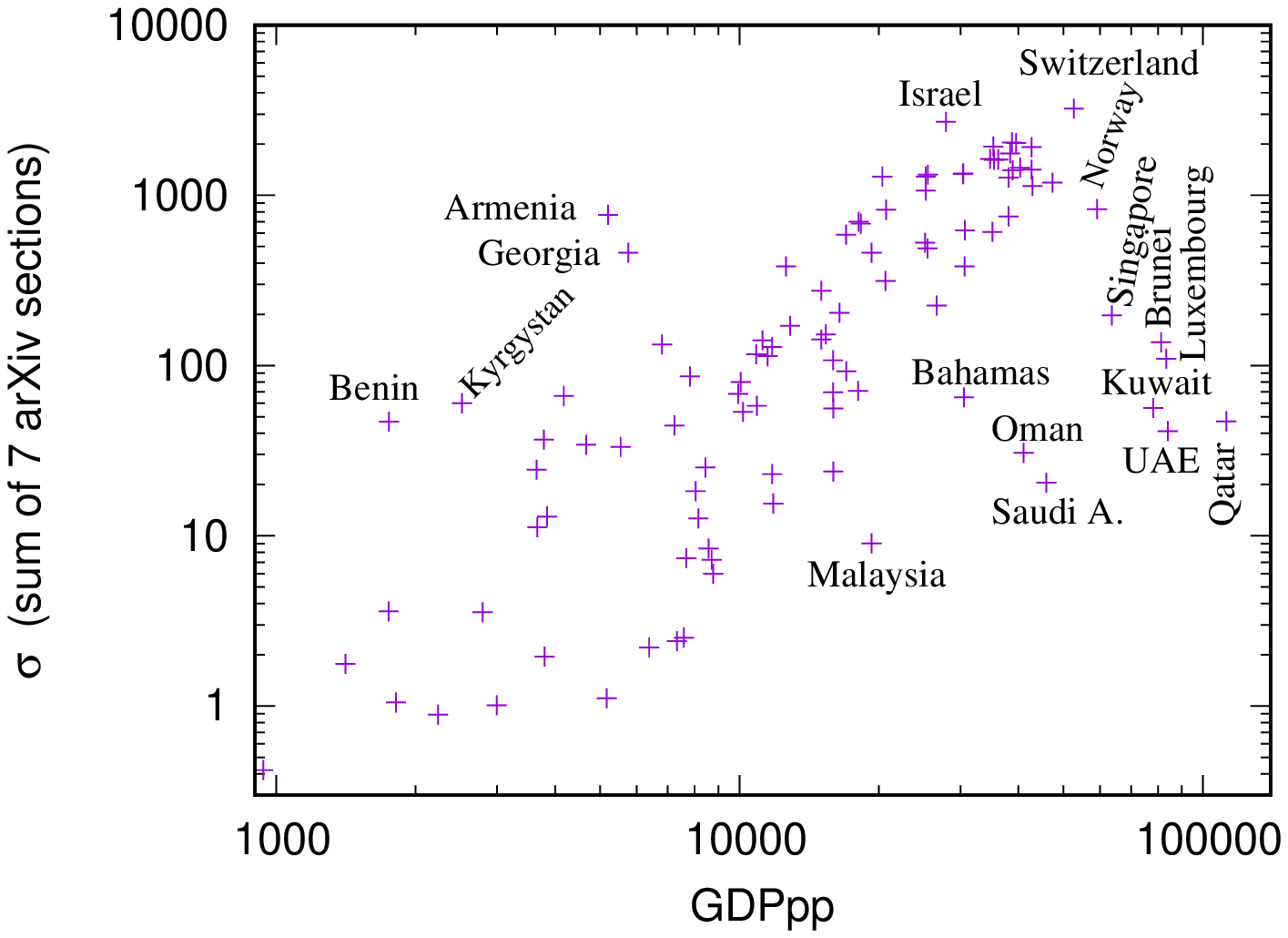}
  \hspace*{-5mm}
  \includegraphics[angle=270,width=.51\linewidth]{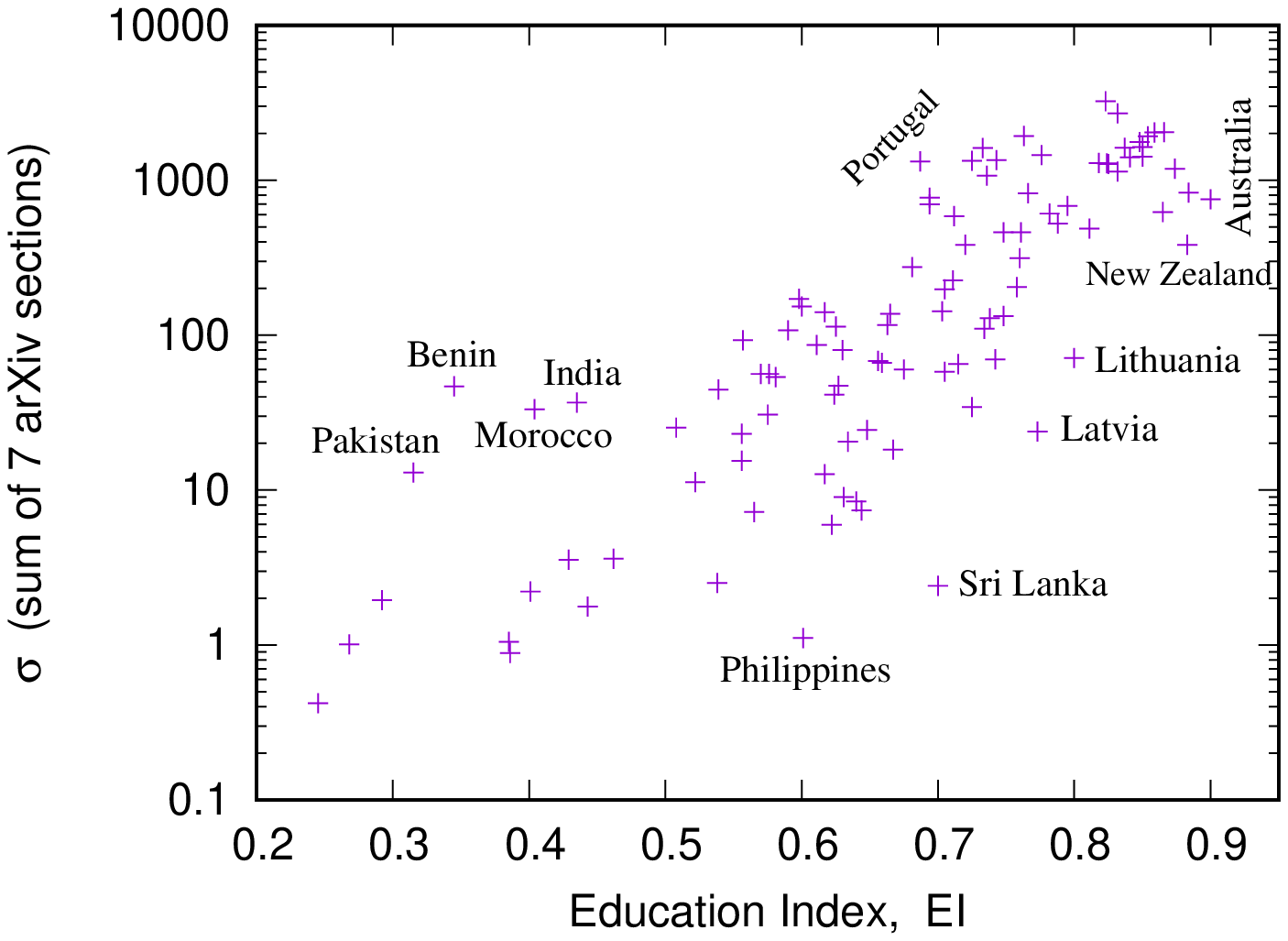} 
\end{center}
\vspace*{-5mm}
\caption{Like Figure \ref{scatterheplat}, but for the
sum over 7 {\tt arXiv} sections.
The plots capture the 101 nations with $H > 5$, the leading 45
can be identified from Table \ref{7arxivrankingint}.
The monotonic trend is somewhat clearer than in Figure \ref{scatterheplat},
which is lattice-specific.
Again we indicate some nations significantly off the dominant trend.}
\label{scatter7arxiv}
\end{figure}

\begin{table}
\centering
{\small
\vspace*{-7mm}
\begin{tabular}{|l c||r|r||r|r|r||r|r|}
  \hline
& e-rank & GDPpp & EI~ & $e$~~~ & $p$~~~ & $c$~~~ & $\Sigma$/GDP & $\sigma$~~~~~ \\
\hline
~1. Cyprus      &  1 & 30386 & 0.725 & 383.32 & 190.18 & 8665.6 & {\bf 44.17} & {\bf 1006.78} \\
~2. Switzerland &  8 & 52698 & 0.823 &  60.13 &  30.40 & 2232.5 & {\bf  3.78} & {\bf  202.16} \\
~3. Ireland     &  3 & 42903 & 0.832 &  70.27 &  28.16 & 2061.1 & {\bf  4.56} & {\bf  201.48} \\
~4. Denmark     &  5 & 42671 & 0.854 &  59.22 &  32.00 & 1873.8 & {\bf  4.31} & {\bf  184.02} \\
~5. Germany     &  4 & 38719 & 0.866 &  55.60 &  25.90 & 1766.1 & {\bf  4.36} & {\bf  169.81} \\
~6. Austria     &  7 & 40348 & 0.776 &  64.31 &  30.59 & 1240.4 & {\bf  3.90} & {\bf  156.92} \\
~7. Hungary     &  2 & 20738 & 0.766 &  29.04 &  15.02 & 1872.0 & {\bf  6.64} & {\bf  137.66} \\
~8. UK          &  9 & 34770 & 0.850 &  32.99 &  15.47 & 1097.4 & {\bf  2.97} & {\bf  103.33} \\
~9. Slovenia    &  6 & 25235 & 0.824 &  33.01 &  16.26 & 1071.1 & {\bf  4.08} & {\bf  102.83} \\
10. Finland     & 10 & 36205 & 0.837 &  31.18 &  15.30 &  842.7 & {\bf  2.45} & {\bf   88.61} \\
11. Italy       & 12 & 35411 & 0.733 &  30.19 &  15.00 &  705.6 & {\bf  2.28} & {\bf   80.47} \\
12. Australia   & 14 & 38067 & 0.900 &  24.95 &  13.25 &  752.8 & {\bf  1.79} & {\bf   75.84} \\
13. Israel      & 11 & 27906 & 0.832 &  22.58 &  14.55 &  700.3 & {\bf  2.41} & {\bf   72.14} \\
14. France      & 15 & 35303 & 0.763 &  17.52 &   8.42 &  688.8 & {\bf  1.66} & {\bf   60.80} \\
15. USA         & 19 & 47324 & 0.874 &  18.59 &   8.72 &  596.9 & {\bf  1.20} & {\bf   57.15} \\
~~~~ European
Union           &    & 31966 & 0.792 &  18.22 &   8.62 &  503.5 & {\bf  1.62} & {\bf   52.01} \\
16. Japan       & 17 & 35125 & 0.782 &  18.96 &   8.45 &  436.8 & {\bf  1.41} & {\bf   49.25} \\
17. Spain       & 16 & 30406 & 0.743 &  16.24 &   8.48 &  442.8 & {\bf  1.54} & {\bf   46.86} \\
18. Slovakia    & 13 & 20657 & 0.760 &  15.19 &   7.78 &  435.9 & {\bf  2.17} & {\bf   44.76} \\
19. Netherlands & 24 & 42689 & 0.850 &  12.48 &   6.02 &  279.3 & {\bf  0.76} & {\bf   32.46} \\
20. Portugal    & 20 & 25515 & 0.687 &  13.33 &   5.31 &  217.5 & {\bf  1.16} & {\bf   29.52} \\
21. Canada      & 25 & 38828 & 0.841 &  10.01 &   5.05 &  258.4 & {\bf  0.72} & {\bf   27.98} \\
22. Sweden      & 26 & 39526 & 0.859 &  10.77 &   5.44 &  235.3 & {\bf  0.70} & {\bf   27.98} \\
23. Taiwan      & 23 & 30654 & 0.865 &   8.70 &   4.44 &  243.7 & {\bf  0.81} & {\bf   25.33} \\
24. Greece      & 22 & 25240 & 0.736 &   8.84 &   4.51 &  161.7 & {\bf  0.85} & {\bf   21.43} \\
25. Belgium     & 31 & 38398 & 0.848 &   5.07 &   2.63 &  168.1 & {\bf  0.42} & {\bf   16.10} \\
26. Iceland     & 34 & 38087 & 0.825 &   3.33 &   3.33 &  143.3 & {\bf  0.36} & {\bf   13.83} \\
27. South Korea & 30 & 25473 & 0.811 &   4.93 &   1.70 &   92.1 & {\bf  0.44} & {\bf   11.24} \\
28. Poland      & 27 & 18267 & 0.795 &   4.72 &   2.14 &   83.3 & {\bf  0.61} & {\bf   11.02} \\
29. New Zealand & 35 & 30601 & 0.883 &   2.66 &   1.93 &  101.4 & {\bf  0.31} & {\bf    9.66} \\
30. Russia      & 32 & 19274 & 0.748 &   3.27 &   1.73 &   59.2 & {\bf  0.41} & {\bf    7.96} \\
31. Albania     & 21 &  7683 & 0.644 &   4.96 &   0.66 &   35.4 & {\bf  0.98} & {\bf    7.39} \\
32. Georgia     & 18 &  5752 & 0.761 &   2.52 &   1.60 &   54.0 & {\bf  1.32} & {\bf    6.82} \\
33. Norway      & 43 & 59066 & 0.884 &   2.96 &   1.27 &   17.6 & {\bf  0.09} & {\bf    5.11} \\
34. Singapore   & 46 & 63641 & 0.705 &   2.22 &   1.11 &   25.0 & {\bf  0.07} & {\bf    4.58} \\
35. Belarus     & 33 & 11763 & 0.738 &   0.93 &   0.62 &   54.4 & {\bf  0.36} & {\bf    4.27} \\
36. Uruguay     & 36 & 15032 & 0.681 &   2.11 &   1.51 &    8.1 & {\bf  0.27} & {\bf    4.03} \\
37. Croatia     & 37 & 18091 & 0.694 &   2.05 &   0.91 &   13.9 & {\bf  0.20} & {\bf    3.65} \\ 
38. Armenia     & 28 &  5207 & 0.694 &   2.30 &   0.66 &    1.6 & {\bf  0.60} & {\bf    3.05} \\   
39. Ukraine     & 29 &  6806 & 0.748 &   1.79 &   0.72 &    8.8 & {\bf  0.44} & {\bf    2.95} \\
40. Brazil      & 39 & 12856 & 0.598 &   0.63 &   0.31 &   21.4 & {\bf  0.16} & {\bf    2.02} \\
41. Czech Rep.  & 48 & 25149 & 0.788 &   0.48 &   0.48 &    6.3 & {\bf  0.05} & {\bf    1.27} \\
42. Mexico      & 45 & 15910 & 0.590 &   0.54 &   0.24 &    7.1 & {\bf  0.07} & {\bf    1.14} \\
43. Jordan      & 40 &  8043 & 0.667 &   0.46 &   0.46 &    2.6 & {\bf  0.13} & {\bf    1.06} \\ 
44. Turkey      & 47 & 16994 & 0.557 &   0.37 &   0.28 &    5.1 & {\bf  0.05} & {\bf    0.90} \\
45. China       & 41 &  7231 & 0.539 &   0.31 &   0.17 &    7.3 & {\bf  0.12} & {\bf    0.85} \\
\hline
\end{tabular}
}
\caption{Intensive {\tt hep-lat} statistics,
  with a ranking according to $\sigma = \Sigma/{\rm pop}$,
  for all nations with $\sigma > 0.8$.
  We add further intensive quantities: GDPpp, EI,
  $(e,p,c) := (E,P,C) /{\rm pop}$.
  We also display the economic rank, e-rank, based on
  $\Sigma/{\rm GDP} \propto \sigma/{\rm GPDpp}$.
  In that regard, India (38), Bangladesh (42) and Yemen (44)
  enter the top 45.}
\label{heplatrankingint}
\end{table}

\begin{table}
\centering
{\small  
\vspace*{-7mm}
\begin{tabular}{|l c||r|r|r||r|r|}
  \hline
& e-rank & $e$~~~ & $p$~~~ & $c$~~~ & $\Sigma$/GDP & $\sigma$~~~~~ \\
\hline
1.~~ Switzerland       &  5 & 760.78 & 549.55 & 38484.1 & {\bf  60.84} & {\bf 3234.53} \\
2.~~ Israel            &  2 & 770.45 & 597.42 & 26678.9 & {\bf  90.16} & {\bf 2701.83} \\ 
3.~~ Germany           &  8 & 556.15 & 436.80 & 21062.6 & {\bf  52.55} & {\bf 2046.09} \\
4.~~ Sweden            & 11 & 541.20 & 408.43 & 21582.7 & {\bf  50.95} & {\bf 2028.76} \\
5.~~ France            &  7 & 468.31 & 345.36 & 22404.5 & {\bf  52.89} & {\bf 1933.89} \\
6.~~ Denmark           & 16 & 552.87 & 413.46 & 19004.5 & {\bf  44.70} & {\bf 1916.56} \\  
7.~~ Belgium           & 13 & 464.85 & 356.31 & 18868.3 & {\bf  45.93} & {\bf 1764.58} \\
8.~~ United Kingdom    & 12 & 426.24 & 324.50 & 17749.2 & {\bf  47.09} & {\bf 1638.20} \\
9.~~ Finland           & 15 & 464.80 & 343.50 & 16268.2 & {\bf  44.79} & {\bf 1621.71} \\
10.~ Italy             & 14 & 462.38 & 399.94 & 16245.2 & {\bf  45.68} & {\bf 1614.58} \\
11.~ Austria           & 27 & 467.86 & 308.66 & 13553.1 & {\bf  30.09} & {\bf 1454.18} \\
12.~ Netherlands       & 24 & 346.27 & 264.71 & 16151.2 & {\bf  33.14} & {\bf 1418.55} \\
13.~ Canada            & 22 & 380.91 & 298.71 & 14475.0 & {\bf  35.89} & {\bf 1403.37} \\
14.~ Spain             & 17 & 377.60 & 280.30 & 13784.6 & {\bf  44.28} & {\bf 1347.12} \\
15.~ Cyprus            &  6 & 489.74 & 267.04 & 11504.6 & {\bf  58.43} & {\bf 1332.02} \\
16.~ Portugal          &  9 & 402.44 & 298.21 & 12450.9 & {\bf  51.88} & {\bf 1323.19} \\
17.~ Estonia           &  4 & 310.98 & 245.85 & 14675.1 & {\bf  64.03} & {\bf 1290.59} \\
18.~ Slovenia          & 10 & 353.76 & 238.47 & 13903.5 & {\bf  51.14} & {\bf 1287.40} \\
19.~ Iceland           & 25 & 476.67 & 380.00 & 8336.7  & {\bf  33.02} & {\bf 1273.50} \\
20.~ United States     & 30 & 290.89 & 215.52 & 13614.6 & {\bf  24.86} & {\bf 1187.14} \\
21.~ Ireland           & 29 & 355.19 & 259.41 & 10419.3 & {\bf  25.69} & {\bf 1135.56} \\
22.~ Greece            & 18 & 322.66 & 252.97 &  9874.5 & {\bf  42.41} & {\bf 1069.35} \\
~~~ European Union     &    & 297.91 & 215.77 & 10022.5 & {\bf  31.59} & {\bf 1014.81} \\
23.~ Norway            & 42 & 262.81 & 195.31 &  7447.9 & {\bf  14.00} & {\bf  830.50} \\
24.~ Hungary           & 19 & 226.14 & 157.72 &  8785.9 & {\bf  39.70} & {\bf  823.16} \\
25.~ Armenia           &  1 & 274.94 & 209.09 &  5706.5 & {\bf 150.46} & {\bf  769.35} \\ 
26.~ Australia         & 39 & 221.73 & 170.55 &  7212.3 & {\bf  17.72} & {\bf  752.90} \\
27.~ Croatia           & 20 & 235.43 & 176.97 &  5735.7 & {\bf  39.12} & {\bf  699.18} \\
28.~ Poland            & 21 & 214.02 & 147.64 &  6393.5 & {\bf  37.49} & {\bf  681.34} \\
29.~ Taiwan            & 34 & 182.44 & 141.43 &  5959.0 & {\bf  19.95} & {\bf  621.82} \\
30.~ Japan             & 40 & 192.89 & 145.94 &  5409.8 & {\bf  17.44} & {\bf  609.32} \\
31.~ Chile             & 23 & 190.64 & 158.42 &  4763.3 & {\bf 33.86}  & {\bf  587.22} \\  
32.~ Czech Rep.        & 33 & 201.85 & 143.87 &  3618.6 & {\bf 21.02}  & {\bf  526.65} \\
33.~ South Korea       & 36 & 158.01 & 126.29 &  4080.2 & {\bf 18.97}  & {\bf  488.31} \\
34.~ Russia            & 31 & 154.18 & 104.90 &  4029.9 & {\bf 23.74}  & {\bf  460.58} \\
35.~ Georgia           &  3 & 158.70 & 117.25 &  3671.7 & {\bf 88.78}  & {\bf  459.54} \\
36.~ New Zealand       & 46 & 119.54 &  87.42 &  3520.1 & {\bf 12.42}  & {\bf  382.98} \\
37.~ Bulgaria          & 26 & 134.16 &  88.24 &  3205.6 & {\bf 30.90}  & {\bf  382.68} \\
38.~ Slovakia          & 41 & 118.89 &  72.96 &  2439.4 & {\bf 15.20}  & {\bf  313.82} \\
39.~ Uruguay           & 38 &  92.61 &  76.32 &  2119.0 & {\bf 18.14}  & {\bf  274.89} \\
40.~ Malta             & 54 &  82.50 &  72.50 &  1405.0 & {\bf  8.18}  & {\bf  225.25} \\
41.~ Argentina         & 44 &  66.17 &  55.97 &  1641.4 & {\bf 12.29}  & {\bf  204.21} \\
42.~ Singapore         & 69 &  82.24 &  62.29 &  1055.8 & {\bf  2.99}  & {\bf  197.32} \\
43.~ Brazil            & 43 &  62.14 &  48.59 &  1207.4 & {\bf 13.16}  & {\bf  171.11} \\
44.~ Iran              & 50 &  58.02 &  48.63 &   923.5 & {\bf  9.81}  & {\bf  152.83} \\
45.~ Romania           & 51 &  58.77 &  42.72 &   823.5 & {\bf  9.69}  & {\bf  142.67} \\
\hline
\end{tabular}
}
\caption{Like Table \ref{heplatrankingint}, but
joint statistics for the 7 {\tt arXiv} sections.
The e-rank strongly deviates from the $\sigma$-rank for
Georgia, Armenia, Estonia, Bulgaria (up), and for
Singapore, Norway, Austria, Malta, Australia, Netherlands (down).
According to the e-rank, Benin (28), Kyrgyzstan (32), Ukraine (35),
Moldova (37) and Lebanon (45) are among the top 45.}
\label{7arxivrankingint}
\end{table}

\end{document}